\documentclass[11pt]{article}

\usepackage[margin=1in]{geometry}
\usepackage{amsmath}
\usepackage{amssymb}
\usepackage{booktabs}
\usepackage{graphicx}
\usepackage{listings}
\usepackage{microtype}
\usepackage{url}
\usepackage[hidelinks]{hyperref}

\title{Retrospective Evaluation of Bluetooth RSSI Fingerprinting for Bin Localization on an Industrial Shop Floor}
\author{Saif Lakhani\\
University of California, Davis\\
\texttt{slakhani@ucdavis.edu}\thanks{The author holds an M.S. in Computer Science from the University of California, Davis. The 2019 data collection and prototype described here were carried out at an industrial site before the author's enrollment, and were not sponsored, funded, or supervised by the University of California, Davis. The affiliation reflects the author's academic association only. This work is likewise unrelated to the author's employment at Amazon: the 2019 system predates that employment}}
\date{October 2026}

\begin{document}
\maketitle

\begin{abstract}
This paper evaluates an archival Bluetooth Low Energy received signal strength indicator (RSSI) dataset collected for bin localization on an industrial shop floor. The installation used 24 fixed relays positioned on walls and Bluetooth beacons attached to bins, with each relay applying a Kalman filter to its own readings in firmware before publishing them. The primary survey produced 19,034 usable two-minute fingerprint samples across 90 known positions, with 23 relay measurements available as model features. We compare calibrated path-loss multilateration with two radio-map estimators: distance-weighted $k$-nearest neighbours and Random Forest regression. A random 70/30 row split measures recognition at surveyed positions, while a coordinate-grouped split reserves 27 entire positions for testing and measures interpolation to unseen positions. Random Forest achieved a median error of 0.66~m for surveyed positions, with 77.9\% of estimates within 3~m and 96.3\% within 10~m. For unseen positions, median error increased to 9.42~m and 55.5\% of estimates were within 10~m. Path-loss multilateration produced a 19.41~m median error on unseen positions, close to the 19.62~m mean-position baseline. The results support RSSI fingerprinting for a fixed inventory of surveyed bin positions, but not precise continuous localization between survey points. Git history dates the original implementation to September 7, 2019. The evaluation code in this paper was written in October 2026 and is reported separately from the seven-year-old prototype.
\end{abstract}


\section{Introduction}
Global Navigation Satellite Systems do not provide dependable positioning inside industrial buildings. Indoor asset tracking therefore relies on local radio infrastructure, inertial sensing, visual markers, or combinations of these systems \cite{zafari2019}. Received signal strength is inexpensive because standard Bluetooth radios expose it without specialized ranging hardware. Its weakness is equally familiar: reflections, obstruction, antenna orientation, and receiver differences can change RSSI without a corresponding change in distance.

This study revisits an industrial tracking prototype whose purpose was to locate movable bins. Bluetooth beacons were attached to the bins, while fixed relays listened from the shop-floor walls. Each position produced a vector of relay RSSI readings. The original project explored regression, classification, and trilateration, but it did not leave a reliable evaluation of coordinate error.

The present work asks two separate questions. First, can a radio map recognize a new reading from a position included in the survey? Second, can the same RSSI measurements infer the coordinate of a position excluded from training? The distinction prevents accuracy at repeated survey points from being presented as continuous indoor localization.

The contributions are:
\begin{enumerate}
    \item a reconstruction of an archival industrial BLE fingerprint dataset with explicit relay and bin geometry;
    \item a test protocol separating surveyed-position recognition from unseen-position interpolation;
    \item a comparison of $k$-nearest neighbours, Random Forest regression, calibrated path-loss multilateration, and a mean-position baseline; and
    \item an account of the limits imposed by sparse relay reception and temporal changes in the recorded radio map.
\end{enumerate}

\section{Related work}
RADAR established radio-map fingerprinting as an approach to indoor positioning by comparing observed signal vectors with measurements collected at known locations \cite{bahl2000}. Horus later modelled the distribution of received signal at each location and used probabilistic inference to improve WLAN positioning \cite{youssef2005}. These systems share a useful assumption: a location may be identifiable from the joint pattern across receivers even when any one RSSI value is a poor distance estimate.

Bluetooth Low Energy provides a lower-power radio source for the same family of methods. Faragher and Harle studied BLE fingerprinting with 19 beacons over an approximately 600~m$^2$ testbed and documented both positioning accuracy and susceptibility to fast fading \cite{faragher2015}. Broader reviews describe RSSI fingerprinting and range-based localization as distinct approaches, with different infrastructure, calibration, and environmental assumptions \cite{zafari2019}.

The deployment studied here reverses the common smartphone arrangement. The beacons move with, or remain attached to, bins, while the relays are fixed receivers. The learning problem is unchanged: map a vector of received signal strengths to a known coordinate.

\section{Archival provenance and scope}
The \href{https://github.com/saiflakhani/Sigma-Tracking}{archival GitHub repository} records its \href{https://github.com/saiflakhani/Sigma-Tracking/commit/87f6aef55f48f95d486f82f1082bda8827579f3c}{first code commit} on September 7, 2019. That commit added the collector, preprocessing scripts, localization experiments, coordinate files, and RSSI datasets. The only subsequent commit was recorded one minute later and changed repository metadata. Thus, according to the GitHub timestamps, the original implementation is at least seven years old as of October 2026. The measurements themselves are dated January 17 and March 1, 2019.

This distinction affects the interpretation of the results. The original files document an unfinished 2019 prototype. Several scripts reference missing intermediate files, the pivoted regression selects only the Y coordinate as its target, and some scripts treat zero as an observed signal rather than an absent reading. The original repository also contains a Python 3.7 environment and library versions from that period.

\section{System geometry and data}
\subsection{Deployment geometry}
The coordinate files describe 24 fixed relays and 202 beacon identifiers. Relays 1 through 10 follow the upper boundary, relays 11 and 12 lie on the right boundary, relays 13 through 21 follow the lower boundary, and relays 23 and 24 lie on the left boundary. Relay 22 appears inside the outer perimeter in the coordinate data. Since the repository does not include an internal wall plan, no wall segment is inferred for that relay.

\begin{figure*}[t]
    \centering
    \includegraphics[width=\textwidth]{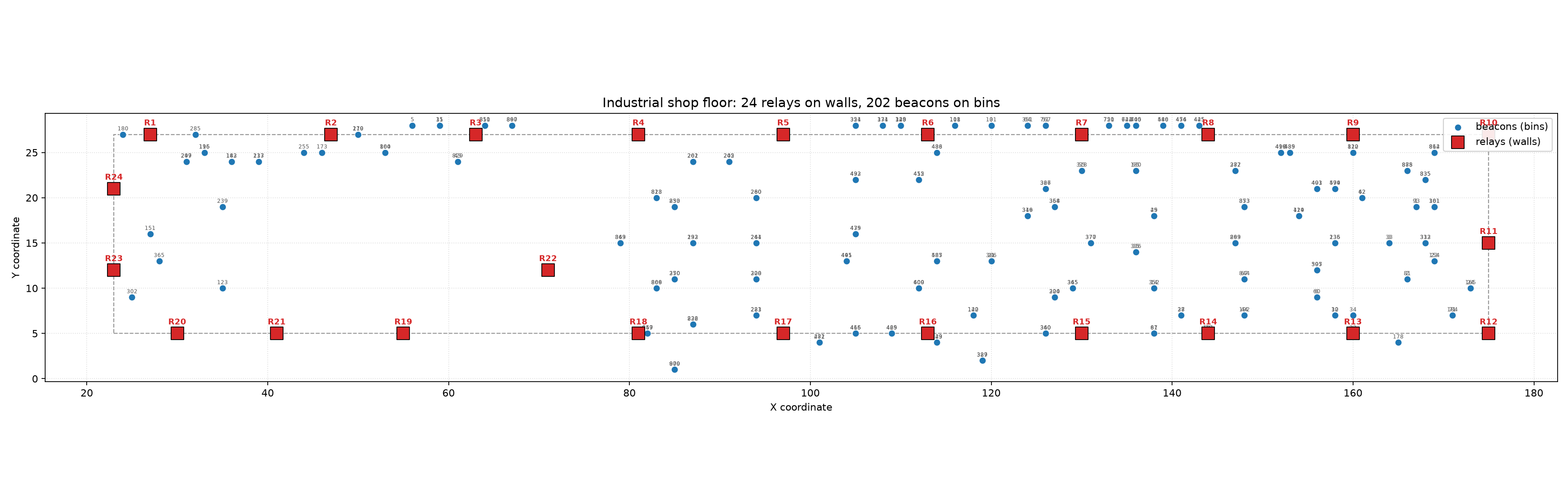}
    \caption{Recorded shop-floor geometry. Red squares are fixed wall relays and blue points are beacon identifiers attached to bins. The coordinate inventory contains 202 beacon identifiers. The primary evaluation contains 90 usable coordinate positions.}
    \label{fig:floor}
\end{figure*}

Beacons were attached to bins at known coordinates. The observed coordinate span is 149~m along X and 26~m along Y. Figure~\ref{fig:floor} shows fixed relays as red squares and mapped bin beacons as blue points.

\subsection{Fingerprint representation}
The top-level \texttt{gatherdata.csv} contains 20,928 rows. Each row has a timestamp, beacon identifier, relay values, and ground-truth X and Y coordinates. Columns 1 through 21, 23, and 24 provide 23 relay features. Relay 22 has a coordinate but no feature column in this table and was not synthesized.

The historical preprocessing stores the magnitude of RSSI. A value of 87 is therefore stronger than 92. Zero indicates that a relay did not receive the beacon during that time bucket. The analysis converts zero and missing cells to a sentinel of 105, weaker than the weakest retained measurement. Values outside 1 to 100 are treated as invalid. No nonzero value in the top-level dataset failed this range check.\\

Listing~\ref{lst:csv} reproduces the header and first three data rows from the original top-level CSV without changing the values. The leading unnamed field is the row index written by the historical preprocessing script.

\begin{lstlisting}[
    caption={Exact four-line excerpt from the original \texttt{gatherdata.csv}.},
    label={lst:csv}
]
,timeStamp,beaconId,1,2,3,4,5,6,7,8,9,10,11,12,13,14,15,16,17,18,19,20,21,23,24,X-Coordinate,Y-Coordinate
0,2019-01-17 07:50:00+00:00,14,0.0,0.0,0.0,0.0,0.0,0.0,0.0,0.0,0.0,0.0,0.0,0.0,0.0,90.0,0.0,0.0,0.0,0.0,0.0,0.0,0.0,0.0,0.0,138,10
1,2019-01-17 07:52:00+00:00,14,0.0,0.0,0.0,0.0,0.0,0.0,0.0,85.5,0.0,0.0,91.0,0.0,0.0,92.0,87.0,0.0,0.0,0.0,0.0,0.0,0.0,0.0,0.0,138,10
2,2019-01-17 07:54:00+00:00,14,0.0,0.0,0.0,0.0,0.0,0.0,91.5,85.5,0.0,0.0,0.0,0.0,0.0,89.5,0.0,0.0,0.0,0.0,0.0,0.0,0.0,0.0,0.0,138,10
\end{lstlisting}

The January 17 session contains 20,615 rows over 157 two-minute timestamps. Rows heard by fewer than three relays were removed, leaving 19,034 observations across 90 coordinate positions. The median usable observation was heard by five relays. The file also contains repeated signal vectors, which can make a random row split optimistic.

The March 1 data contains 313 rows and is not included in the primary accuracy estimates. Its median observation was heard by two relays, four relays recorded no observations, and only 151 rows at three coordinate positions survived the minimum-relay filter. These records are discussed as a diagnostic rather than a repeatability experiment.

\subsection{Temporal preprocessing as archived}
\label{sec:preproc}
The evaluated tables are not raw radio observations. Each reading arrives already smoothed by the on-relay Kalman filter of Section~\ref{sec:kalman}, and the archived preprocessing script then applies a second, median-based reduction. That second stage is specified here because it determines the noise characteristics of every result that follows. Let $z_{b,j,t}$ denote a relay-reported RSSI magnitude for beacon $b$ at relay $j$ and time $t$. The script first discards weak observations, retaining only $z_{b,j,t} < 95$. It then forms a per-pair median over the whole session,
\begin{equation}
    m_{b,j} = \operatorname{median}_{t}\left\{ z_{b,j,t} \right\},
\end{equation}
and applies a deviation gate that keeps a reading only when it lies within two units of that median,
\begin{equation}
    \mathcal{K}_{b,j} = \left\{ t \;:\; \left| z_{b,j,t} - m_{b,j} \right| \leq 2 \right\}.
    \label{eq:gate}
\end{equation}
Surviving readings are assigned to two-minute buckets, $\tau(t) = \lfloor t \rfloor_{2\min}$, and each bucket is summarized by a second median,
\begin{equation}
    r_{b,j,\tau} = \operatorname{median}\left\{ z_{b,j,t} \;:\; t \in \mathcal{K}_{b,j},\; \tau(t) = \tau \right\}.
    \label{eq:bucket}
\end{equation}
The pivoted value $r_{b,j,\tau}$ is the feature consumed by every estimator in this paper, and an empty bucket yields the absent-relay sentinel described above.

Two consequences deserve emphasis. The gate in Equation~(\ref{eq:gate}) is tight: readings more than two units from the session median for that pair are removed before any model sees them, so the retained distribution is narrower than the physical one and the reported errors are conditioned on that filtering. The double median in Equations~(\ref{eq:gate}) and~(\ref{eq:bucket}) also suppresses short-lived fading, which is the dominant fast-varying error term in BLE fingerprinting \cite{faragher2015}, meaning the evaluation understates the variability a live system would face between survey intervals.

\section{Localization methods}
Let $\mathbf{r}_i \in \mathbb{R}^{23}$ denote the relay fingerprint for observation $i$, and let $\mathbf{p}_i = (x_i,y_i)$ denote its known bin coordinate.

\subsection{Path-loss multilateration}
\label{sec:multilateration}
The range-based method fits a global log-distance relation on training observations:
\begin{equation}
    q = A + B \log_{10}(d),
\end{equation}
where $q$ is RSSI magnitude and $d$ is the relay-to-bin distance in metres. For each test observation, the fitted relation is inverted to estimate one distance per receiving relay. The predicted coordinate minimizes the residual between estimated and geometric distances:
\begin{equation}
    \hat{\mathbf{p}} = \arg\min_{\mathbf{p}} \sum_{j \in \mathcal{H}}
    \left(\lVert \mathbf{p} - \mathbf{a}_j \rVert_2 - \hat{d}_j\right)^2,
\end{equation}
where $\mathcal{H}$ is the set of relays that heard the beacon and $\mathbf{a}_j$ is relay $j$'s known coordinate. Estimates are constrained to the observed floor bounds.

On the unseen-position split, the fitted relation was $q = 81.56 + 5.54\log_{10}(d)$. The shallow fitted slope leaves distance estimates sensitive to small RSSI changes.

\subsection{Radio-map regression}
\label{sec:regression}
Two supervised estimators map the complete fingerprint directly to coordinates. The first is distance-weighted $k$-nearest-neighbour regression with $k=5$. The second is a Random Forest regressor with 200 trees and random seed 7. Both predict X and Y jointly. Beacon identifiers and timestamps are excluded from the feature matrix. No hyperparameter search was performed.

\subsection{On-relay Kalman smoothing}
\label{sec:kalman}
Each relay applied a Kalman filter \cite{kalman1960} to its own RSSI readings before publishing them, so the values in the archived dataset are already smoothed in firmware. The filter ran separately for each beacon the relay could hear, and it tracked one number: that beacon's current signal strength.

The idea is a running estimate corrected by each new reading. Let $\hat{x}_{k-1}$ be the previous estimate and $z_k$ the new raw reading. The filter updates
\begin{equation}
    \hat{x}_k = (1 - K)\,\hat{x}_{k-1} + K\,z_k ,
    \label{eq:kalmanupdate}
\end{equation}
where the gain $K$ lies between 0 and 1 and sets how much the new reading is trusted. A small $K$ keeps the estimate steady and ignores short spikes; a large $K$ follows the raw signal closely. The gain is derived from two variances supplied to the filter: $Q$, how much the true signal is expected to drift between readings, and $R$, how noisy an individual reading is. Roughly, more measurement noise lowers $K$ and more expected drift raises it.

For a stationary beacon the gain settles to a constant after a few readings, so Equation~(\ref{eq:kalmanupdate}) behaves like a weighted moving average of the signal. This is useful for suppressing momentary fades, and it is the reason the filter did not need to be retained in later processing: the median summary described in Section~\ref{sec:preproc} removes the same short-lived spikes, and a median is also less disturbed by one very bad reading than a Gaussian average is. Neither step estimates position. Both only clean the signal that the estimators of Sections~\ref{sec:multilateration} and~\ref{sec:regression} consume.

\paragraph{Reproducibility status.}
The filter was implemented in relay firmware. Its parameters ($Q$ and $R$) and its filtered-versus-raw behaviour therefore cannot be recovered from the released code, and no record of its configuration survives. The practical consequence for this paper is that the dataset is smoothed twice: once on the relay, then again by the median pipeline of Section~\ref{sec:preproc}. Both stages reduce the short-term variability that a live system would encounter, so the error figures reported below should be read as a lower bound on the variability of an unfiltered deployment.

\section{Experimental design}
\subsection{Surveyed-position split}
A random 70/30 row split from January 17 produced 13,323 training observations and 5,711 test observations. All 90 coordinate positions occur in both partitions. This test measures recognition at surveyed positions. It does not measure generalization to an unmeasured coordinate.

\subsection{Unseen-position split}
A grouped 70/30 split assigns complete coordinates to one partition only. The training set contains 63 positions and the test set contains 27 disjoint positions, with 12,958 and 6,076 observations respectively. This test measures interpolation to bin coordinates absent from the radio map. The deterministic split uses random seed 7.

\subsection{Metrics}
For prediction $\hat{\mathbf{p}}_i$, Euclidean position error is:
\begin{equation}
    e_i = \lVert \hat{\mathbf{p}}_i - \mathbf{p}_i \rVert_2.
\end{equation}
We report median, mean, and 90th-percentile error in metres. Threshold accuracy at radius $t$ is:
\begin{equation}
    \mathrm{Acc}(t) = \frac{1}{N}\sum_{i=1}^{N}\mathbb{1}[e_i \leq t],
\end{equation}
for $t \in \{3,5,10\}$ metres. A baseline always predicts the mean training coordinate.

\section{Results}
\label{sec:results}
Table~\ref{tab:results} reports the primary results. At surveyed positions, Random Forest achieved a median error of 0.66~m and mean error of 2.07~m. It placed 77.9\% of observations within 3~m, 86.8\% within 5~m, and 96.3\% within 10~m. The $k$-nearest-neighbour model achieved an effectively zero median because 59.3\% of its test estimates matched the recorded coordinate exactly. Repeated vectors and shared positions make this the more favourable protocol.

When complete positions were held out, Random Forest median error rose to 9.42~m and mean error to 10.09~m. Its 3~m accuracy fell to 5.8\%, while 55.5\% of observations remained within 10~m. The $k$-nearest-neighbour result was similar, with 9.42~m median error and 54.8\% within 10~m.

\begin{figure*}[t]
    \centering
    \includegraphics[width=0.95\textwidth]{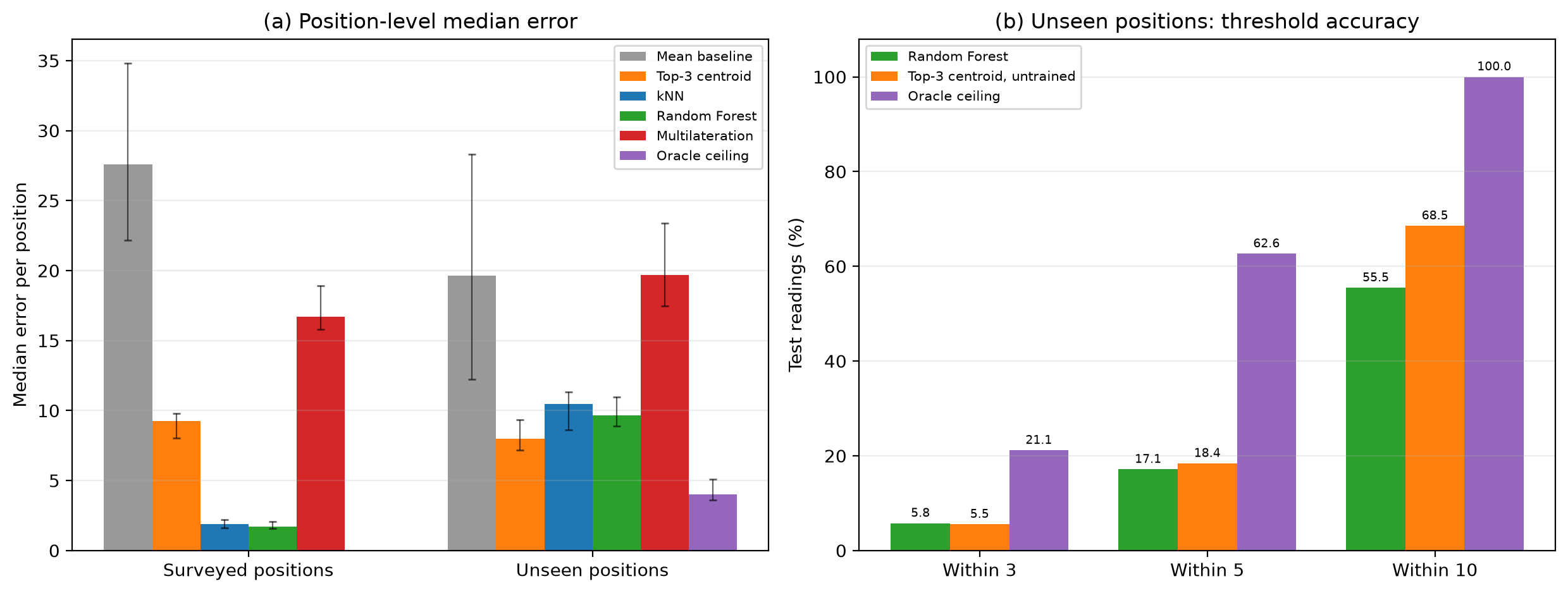}
    \caption{Primary localization results. Panel (a) compares median Euclidean error for surveyed and unseen positions. Panel (b) reports Random Forest threshold accuracy.}
    \label{fig:results}
\end{figure*}

Path-loss multilateration performed poorly in both tests. Its unseen-position median error was 19.41~m, compared with 19.62~m for the mean-position baseline. Only 24.7\% of multilateration estimates were within 10~m. Figure~\ref{fig:results} compares the methods and shows Random Forest threshold accuracy under both splits.

\begin{table*}[t]
\centering
\caption{Localization performance. Threshold columns report the percentage of test observations within the stated radius.}
\label{tab:results}
\begin{tabular}{llrrrrr}
\toprule
Test & Method & Median (m) & Mean (m) & Within 3 m & Within 5 m & Within 10 m \\
\midrule
Surveyed positions & $k$NN & 0.00 & 2.11 & 77.2\% & 84.9\% & 94.2\% \\
Surveyed positions & Random Forest & 0.66 & 2.07 & 77.9\% & 86.8\% & 96.3\% \\
Surveyed positions & Multilateration & 15.03 & 19.11 & 4.2\% & 10.2\% & 30.9\% \\
\midrule
Unseen positions & $k$NN & 9.42 & 10.24 & 6.9\% & 16.9\% & 54.8\% \\
Unseen positions & Random Forest & 9.42 & 10.09 & 5.8\% & 17.1\% & 55.5\% \\
Unseen positions & Multilateration & 19.41 & 21.97 & 3.5\% & 7.4\% & 24.7\% \\
\bottomrule
\end{tabular}
\end{table*}

\section{Discussion}
The random-split result supports a discrete radio map. When a bin occupies a surveyed position and the live fingerprint resembles the training survey, the model can identify that coordinate with useful accuracy. Random Forest produced the best threshold rates, while $k$-nearest neighbours produced more exact coordinate matches.

The unseen-position result does not support precise continuous localization. A 9.42~m median error may separate broad regions of a 149~m-long floor, but it is insufficient for selecting a specific bin unless candidate positions are widely separated. The close agreement between the two learning methods suggests that changing the regressor alone will not resolve this limitation. Additional surveyed positions and more consistent relay coverage are more likely to improve performance.

The multilateration result explains why the range-based prototype struggled. A global RSSI-to-distance equation poorly represents this environment. The fitted signal-distance slope is shallow, and the resulting distance error is amplified when several circles are combined into a coordinate. The radio-map models avoid explicit distance estimation and instead use the joint pattern across relays.

A deployment should also be allowed to abstain. Observations heard by fewer than three relays should return no position. The same applies when a test fingerprint is far from every training fingerprint. Forced coordinates hide infrastructure failures and distribution shifts.

\section{Limitations}
The dataset represents one industrial floor and one primary survey date. Ground-truth coordinates identify bin positions but do not provide an independent measurement of beacon motion within a bin area. Multiple beacon identifiers may share a coordinate. The random row split includes repeated positions and may place repeated fingerprints in both partitions.

\section{Conclusion}
Archival BLE RSSI measurements can locate bins at positions represented in a shop-floor radio map. Random Forest placed 77.9\% of held-out readings within 3~m and 96.3\% within 10~m when the positions were included in training. Accuracy fell sharply at unseen positions, where median error was 9.42~m and only 55.5\% of readings landed within 10~m. Calibrated multilateration was close to a mean-position baseline.

These results support fingerprinting for a fixed set of surveyed bin locations. They do not support precise continuous coordinates from the available RSSI measurements. The next experiment should repeat a dense subset of the survey on a later date, verify all relays before collection, and evaluate grouped cross-validation across positions and collection sessions.

\end{document}